\documentclass[journal]{IEEEtran}
\ifCLASSINFOpdf
\else
\fi
\usepackage{subfigure}
\usepackage{color}
\usepackage{graphicx}
\usepackage{amssymb}
\usepackage{amsmath}
\usepackage{tikz,xcolor,hyperref}
\definecolor{lime}{HTML}{A6CE39}
\DeclareRobustCommand{\orcidicon}{%
	\begin{tikzpicture}
		\draw[lime, fill=lime] (0,0) 
		circle [radius=0.16] 
		node[white] {{\fontfamily{qag}\selectfont \tiny ID}}; 
		\draw[white, fill=white] (-0.0625,0.095) 
		circle [radius=0.007];	  
	\end{tikzpicture}
	\hspace{-2mm}}
\foreach \x in {A, ..., Z}{%
	\expandafter\xdef\csname orcid\x\endcsname{\noexpand\href{https://orcid.org/\csname orcidauthor\x\endcsname}{\noexpand\orcidicon}}
}
\begin{document}

\title{Distributed Cooperative Control of BESSs in AC and DC Hybrid Microgrid and Its Energy Internet Paradigm}
\newcommand{\orcidauthorA}{0000-0002-3788-700X}
\newcommand{\orcidauthorB}{0000-0002-3565-4800} 
\newcommand{\orcidauthorC}{0000-0002-1415-4073} 

\author{Yalin Zhang\orcidA{},
        Zhongxin Liu\orcidB{},~\IEEEmembership{Member,~IEEE,}
        and Zengqiang Chen\orcidC{}
\thanks{Manuscript received XX, XX; revised XX, XX;
	accepted XX, XX. This work is supported by the Tianjin Natural Science Foundation of China (Grant No.20JCYBJC01060) , and the National Natural Science Foundation of China (Grant No. 62103203, 61973175) and the General Terminal IC Interdisciplinary Science Center of Nankai University (Corresponding author: Zhongxin Liu.). 
	\par The authors are with the College of Artificial Intelligence, Nankai University, Tianjin 300350, and also with the Tianjin Key Laboratory of Brain Intelligent Rehabilitation, Nankai University, Tianjin 300350, China (e-mail: zhangyl@mail.nankai.edu.cn; lzhx@nankai.edu.cn; chenzq@nankai.edu.cn).}}

\markboth{IEEE TRANSACTIONS ON POWER SYSTEMS, VOL. XX, NO. XX, XX XX}%
{Zhang \MakeLowercase{\textit{et al.}}: Distributed Cooperative Control of BESSs in AC and DC Hybrid Microgrid and Its Energy Internet Paradigm}

\maketitle

\begin{abstract}
	An AC and DC hybrid microgrid, which inherits advantages of AC and DC microgrids and discards some disadvantages, is considered to be the most promising power network structure and gradually applied in the community. Usually, the AC subgrid and the DC subgrid are interconnected by Bidirectional Interlink Power Converters (BILPCs). Besides, in view of the different droop characteristics of AC subgrid and the DC subgrid, it is necessary to design a suitable distributed secondary controller for an AC and DC hybrid microgrid. Accordingly, this paper proposes a flexible and scalable distributed control framework for an AC and DC hybrid battery energy storage system (ADHB) with BILPCs in an Energy Internet (EI) paradigm. An ADHB governed by multi agent systems via a cloud server can reach the State-of-Charge balance, proportional power sharing, frequency and voltage restoration. The proposed control framework provides the group play-and-plug by adding or removing an inter-MASs interaction link. For a single BILPC in an ADHB, active/reactive power, frequency and voltage are adjusted by an AC BESS. For the parallel BILPCs in EI, a decentralized secondary control scheme is proposed. Communication delay issues and stability are analysed. Then, the relevant simulation results verify the correctness of the proposed scheme.
\end{abstract}


\IEEEpeerreviewmaketitle

\section{Introduction}

\IEEEPARstart{I}{n} order to meet the challenge brought by the hybrid energy utilization, the concept of Energy Internet (EI) comes into being, which is a typical application and focuses on developing and utilizing various forms of energy sources, enhancing system security and managing distributed energy sources flexibly [1]. Smart grid, as a key concept, is aimed at providing a reliable and efficient power system, where real power system, communication technology and information technology are involved [2], [3]. Thus, the components and technologies required for an EI can be summarized as microgrids (MGs), a communication network and control algorithms. The successful implementation of an EI involves many disciplines, and the challenges it brings have aroused great interest of many scholars.
\par MG, as a new generation and distribution mode, is able to operate both in islanded and grid-connected mode [4], [5]. Hierarchical control is used to achieve their respective control or optimization objectives [6]. Droop control, as a frequently-used primary controller, is used for autonomous power-sharing among all distributed generators (DGs) without any communication with each other. Secondary control is responsible for the frequency/voltage restoration [7], [8] and accurate power sharing in an islanded MG. Tertiary control, actually an optimization problem, is aimed at realizing the optimal operation of a MG, in which the optimal dispatch values are calculated [9], [10]. Tertiary control is actually a discrete controller, and during each dispatch interval, the deviated active power is handed over to the primary and secondary controllers for processing [11], [12]. Compared with a centralized controller, a MG with a distributed secondary and tertiary controllers is more flexible and scalable. Distributed solutions implemented for a MG are often based on multi-agent systems (MASs), where each agent communicates and exchanges states information with its neighbors over a sparse graph [13].
\par According to different voltage forms of a bus, MGs can be divided into DC and AC MGs. Many scholars and researchers have developed secondary controllers for DC and AC MGs. Most of their research focuses on finite-time control [14], event-triggered control [14], [15], fault tolerant control [16], optimal control [17] and the method of combining data-driven and reinforcement learning [18], [19], etc. These designs are aimed at promoting convergence rate [14], communication efficiency [14], [15], robustness against fault [16], dynamic performance [17] and robustness against uncertainty [18], [19], etc. Distributed control scheme for a large-scale AC MG in an EI paradigm is proposed in [20], where the interaction among MGs is considered.
\par For an AC and DC hybrid microgrid (ADHMG), scholars initially focus on themes such as energy management and network planning [21], [22]. In recent years, some secondary control schemes for a ADHMG are proposed. Although the distributed secondary controller is proposed by authors in [23], the proportional power sharing controller is not designed, and Bidirectional Interlink Power Converters (BILPCs) [24], as an important connecting equipment between DC and AC subgrids, is not paid attention to. After that, a more comprehensive scheme appears in [25], whose authors give a secondary scheme based on the nonlinear exponential control. However, as a common form, the control scheme for multi-parallel BILPCs is not given. In [26], a decentralized sliding mode control scheme is designed to achieve  power sharing, and the design method of the reference signal for each controller is given. Unfortunately, there seems to be no suitable distributed scheme for power sharing among paralleled BILPCs. In an ADHMG, the power sharing of multiple parallel BILPCs can improve the security of system operation.
\par From the existing literature, two important issues are rarely involved. The first one is that there is rarely a scalable and flexible solution for large-scale multiple ADHMG clusters. As more and more ADHMG clusters held by different stakeholders are interconnected, the number of controllable units increases. With the support of the EI, a distributed control framework needs to be redesigned to meet the flexible cluster-based plug-and-play. Second, in the existing hierarchical control mode, the secondary controller and the ADHMG entity need to be located at the same position. In fact, the distributed secondary controller can neither be located at the same position as DG, nor can it have a dedicated communication network. In view of this, it is necessary to use the Internet to remotely control the ADHMG. Besides, most of the existing secondary control protocols mostly use the measured values of frequency and voltage, which not only requires additional measuring devices, but also suffers easily from measured errors and faults.
\par To provide a reliable solution for the above two issues, this paper attempts to design a distributed remote control scheme for a AC and DC hybrid battery energy storage system (ADHB) in an EI. Each battery energy storage system (BESS) is flexibly controlled by an agent via a cloud server. With the support of a cloud server, the AC subgrid, DC subgrid and BILPC can be remotely controlled. Both the State-of-Charge (SoC) balance and the proportional power sharing can be successfully reached among all BESSs, whether within or between subgrids. The EI paradigm provides a reliable framework for the group plug-and-play, as well as a flexible distributed control scheme for wide area power grids. For multiple parallel BILPCs, a decentralized control framework is proposed, that is, power sharing, voltage and frequency are regulated by the state variables of AC subgrid. The main contributions made in this article are described below.
\\(1) The problem of MASs managing ADHBs in an EI is studied in this paper, which is rarely reported in the previous literature.
\\(2) An EI paradigm of ADHBs is constructed. Considering the intra-group and inter-group interactions of MASs/ADHBs owned by multiple stakeholders, a distributed secondary control framework for group-based plug-and-play is proposed.
\\(3) The remote control framework is proposed. With a cloud service model as the core, a solution for the remote management of ADHBs by MASs in multiple sites is designed here.
\par The rest sections are arranged as follow. Section II introduces an ADHB and control objectives. Section III presents the control scheme for an ADHB and its EI paradigm with a cloud server. Section IV designs some cases to verify the proposed scheme
\par \emph{Notation}: $0$ in each matrix is a zero matrix with appropriate dimension. $I_M$ denotes an $M\times M$ identity matrix. $diag\{A,B\}$ is a diagonal matrix with $A$ and $B$ as diagonal blocks. $col(a,b)$ denotes column vectorization.
\section{Preliminaries}
\subsection{The descriptions of AC and DC droop-controlled heterogenous BESSs}
\begin{figure}
	\centering
	\subfigure[The control structure for an AC BESS and its hierarchical control.]{\includegraphics[width=8cm]{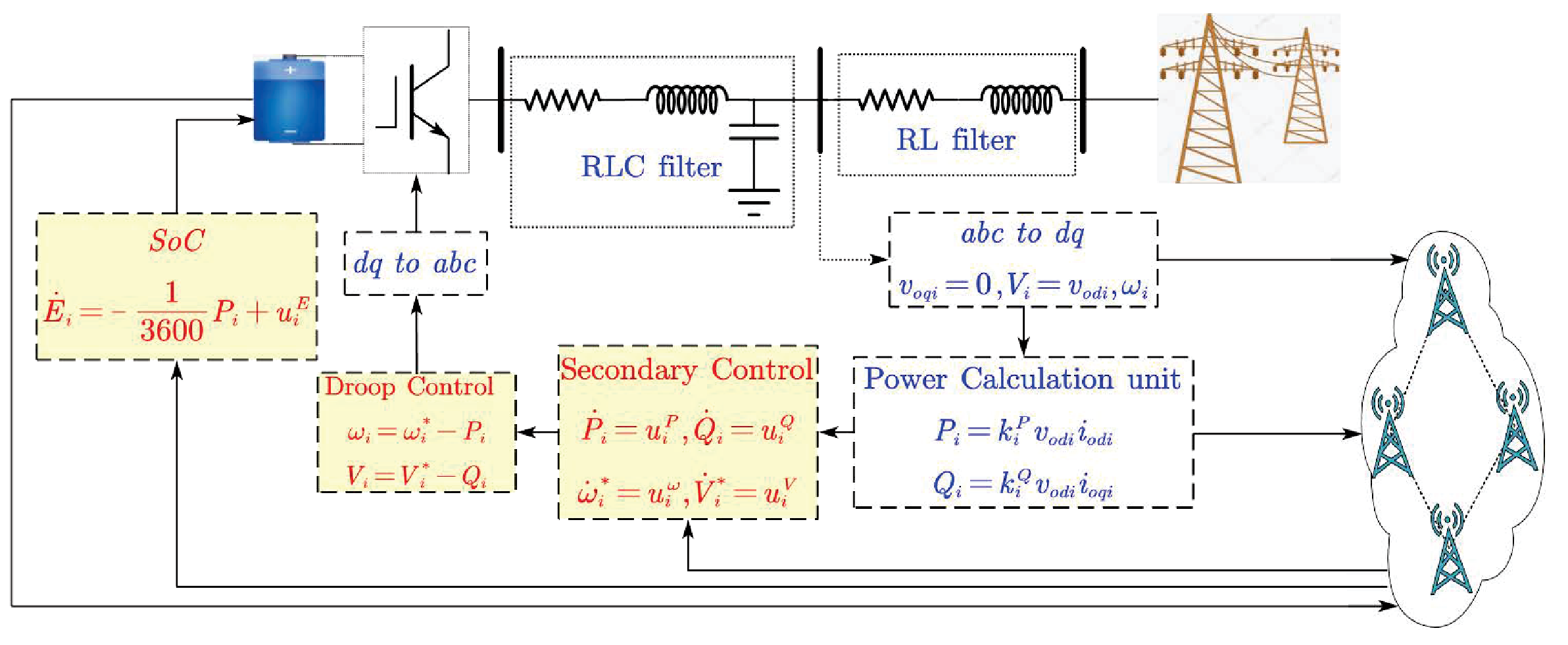}\label{fig1a}}
	\subfigure[The control structure for a DC BESS and its hierarchical control.]{\includegraphics[width=8cm]{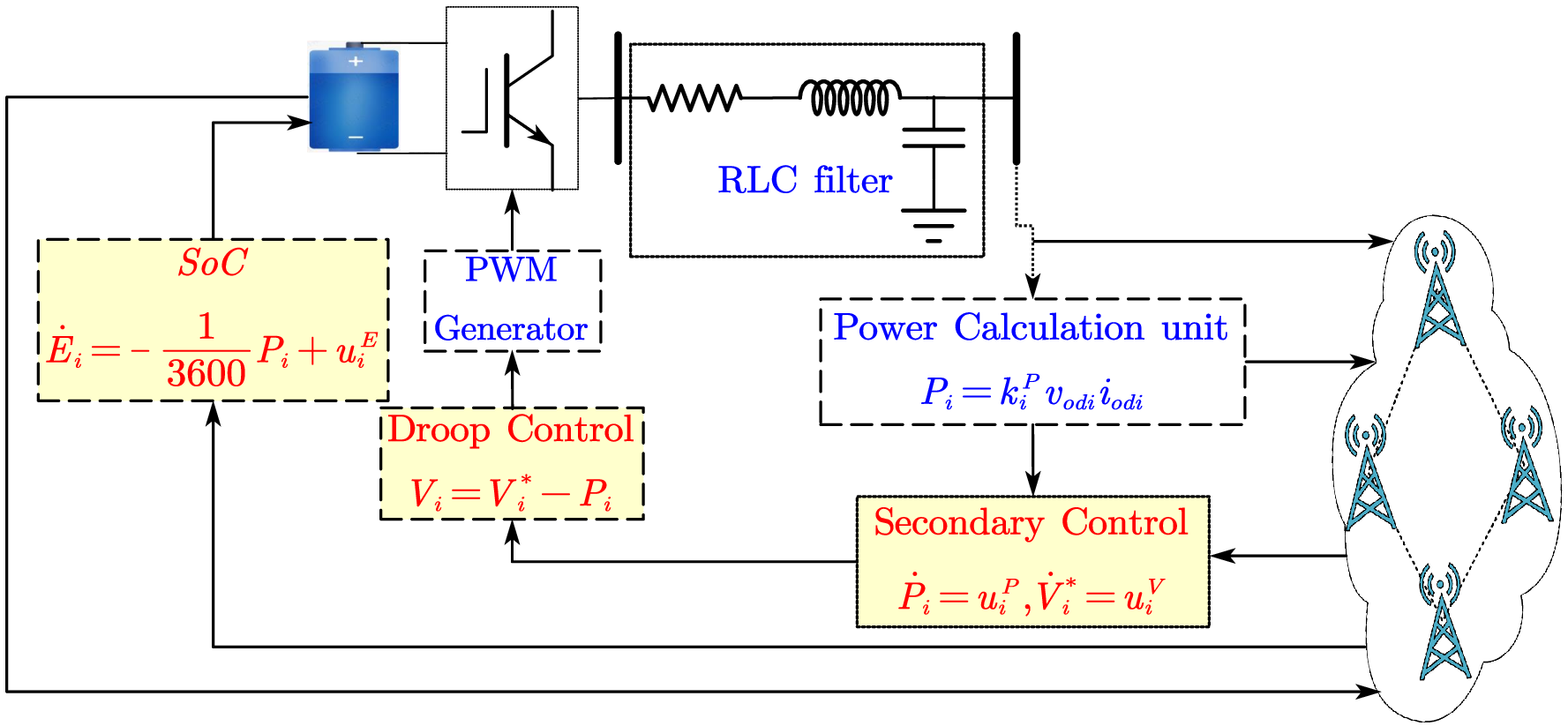}\label{fig1b}}
	\caption{The comparison of AC and DC BESSs and their hierarchical control.}
	\label{fig1}
\end{figure}
\par Generally, the control structure with an inverter/converter as the core component in an AC/DC BESS contains a power calculation unit, a droop controller, a current controller and a voltage controller, as shown in Fig. \ref{fig1}. According to [28], current controller and voltage controller can be reasonably neglected. Thus, the control structure of a BESS can be simplified to a droop controller. 
\par Assuming that there are $n$ heterogenous batteries in BESSs. For simplicity, subsequent variables are simplified by omitting time variable $t$. For battery $i$, the energy storage level $E_i$ is characterized by SoC, whose dynamics can be expressed as
\begin{equation}
	\dot {E_i} =  - \frac{{K_i^E}}{{3600}}{P_i}\label{1}
\end{equation}
where $K_i^E$ reflects the characteristic of battery $i$, ${P_i}$ is the output power of battery $i$.
\par In an AC BESS, droop controllers are organized in (\ref{2a}) and (\ref{2b})
\begin{subequations}
	\begin{equation}
		{\omega _i} = \omega _i^* - K_i^P{P_i}\label{2a}
	\end{equation}
    \begin{equation}
	    {V_i} = V_i^* - K_i^Q{Q_i}\label{2b}
    \end{equation}
\end{subequations}
where ${\omega _i}$ and $\omega _i^*$ are the measured and nominal frequency of BESS $i$, ${V _i}$ and $V _i^*$ are the measured and nominal voltage magnitude of BESS $i$ and referred to the measured and nominal voltage for short, ${P_i}$ and ${Q_i}$ are active and reactive power of BESS $i$, $K_i^P$ and $K_i^Q$ are droop gains.
\par In a DC BESS, it can be approximately considered that there is a droop characteristic between voltage and power as shown in (\ref{3}),
\begin{equation}
	{V_i} = V_i^* - K_i^P{P_i}\label{3}
\end{equation}
where $K_i^P$ is the droop gain. $K_i^P$ is usually closely related to the capacity of battery $i$, which means the heterogeneity. Thus, without loss generality, $K_i^P=K_i^E$.
\subsection{Hierarchical control of an ADHB}
\begin{figure} 
	\includegraphics[width=8cm]  {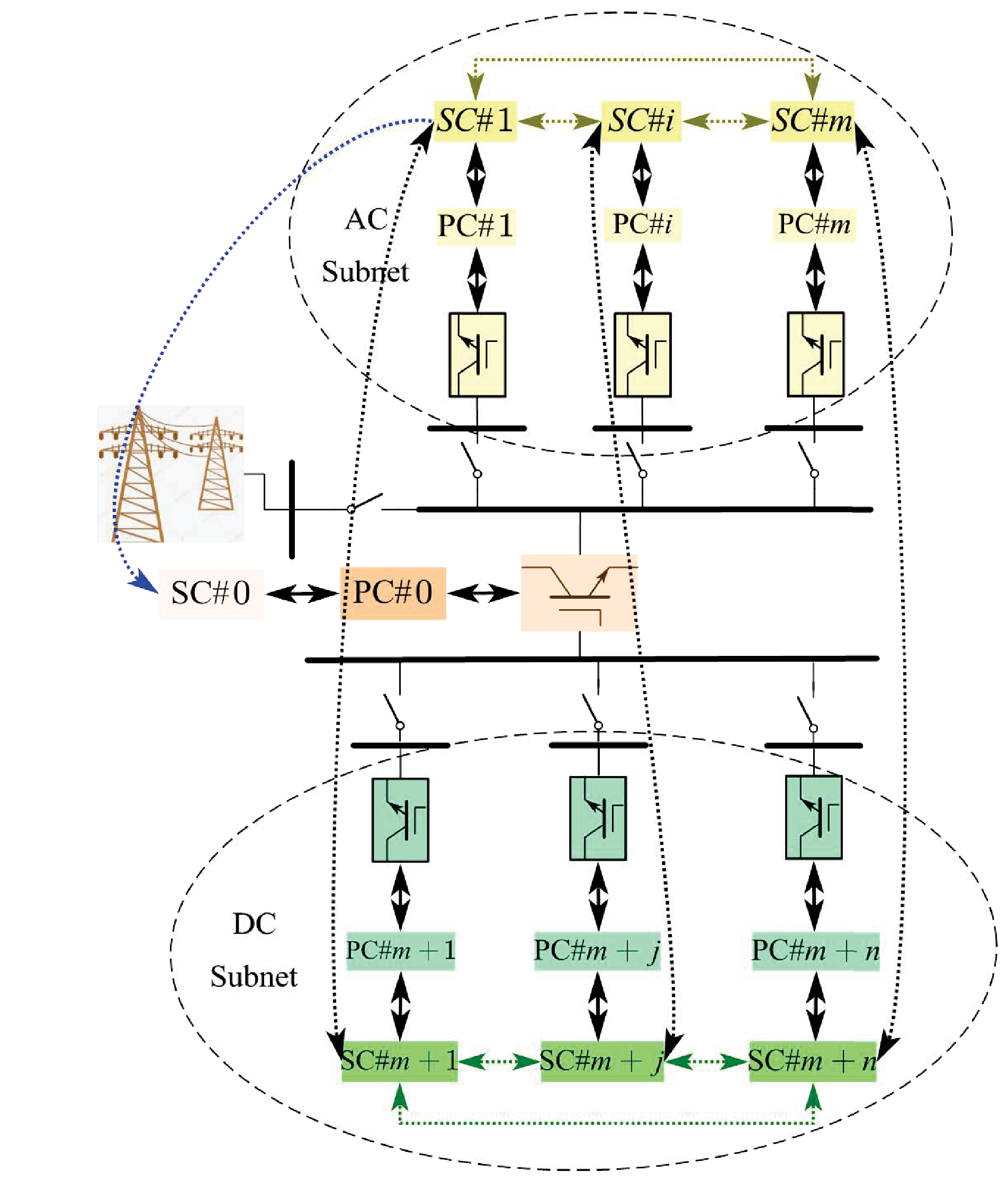} 
	\caption{Hierarchical control mode based on MASs of an ADHB.}\label{fig2}
\end{figure}
\par In an AC/DC BESS, an inverter/converter is adopted and in charge of charging and discharging operation of the battery connected to it. For an ADHB, BILPC, which can realize the inversion and rectification between DC subgrid and AC subgrid, is assumed as a key component for the interconnection between AC subgrid and DC subgrid. The physical connectivity between AC bus and UG is controlled by an electric knife switch. According to the number of BILPCs placed between AC bus and DC bus, there are two structures in the practical application. The one is with a single BILPC and the other is with a set of parallel-connected BILPCs. In this paper, each ADHB is equipped with a single BILPC. Thus, the hierarchical control mode based on MASs of an ADHB is described in detail in Fig.\ref{fig2}.
\par As shown in Fig. \ref{fig2}, a hierarchical control mode is adopted in an ADHB. Droop controller, as the local primary controller (PC), is installed in each BESS. The distributed secondary controller (SC) is designed based on MASs, which is generally installed at a location close to a physical entity. A SC provides the control signals $\{\omega _i^*,V_i^*\}$ for the local controller of AC BESS $i$, $\{ V_j^*\}$ for DC BESS $j$ and $\{\omega _0^*,V_0^*\}$ for the BILPC. The electrical parameters and control signals, i.e., $\{{E_i},{P_i},{Q_i}\}$ and $\{\omega _i^*,V_i^*\}$ of AC BESS $i$ and $\{{E_j},{P_j}\}$ and $\{V_i^*\}$ of DC BESS $j$, are sent their neighbors. Specially, the BILPC acquires the data packets $\{P_i,Q_i,\omega_i^*,V^*_i\}$ from its neighbor AC BESS $i$.
\subsection{The dynamics for an ADHB based on droop control}
\par Consider an ADHB containing $M$ BESSs, $m$ of which are AC and $n$ are DC, and $M=m+n$. Based on the droop control, dynamics of AC and DC BESSs can be expressed as
$$\left\{ \begin{array}{l}
	\dot E=-\frac{1}{3600}P + u^E\\
	\dot P=u^P
\end{array} \right., $$
$$\left\{ \begin{array}{l}
	\dot Q = u^Q \\
	\dot \omega^*=u^\omega
\end{array} \right.,$$
$$\dot V^* = u^V,$$
where $E$ ($u^E$), $P$ ($u^P$), $Q$ ($u^Q$), $\omega^*$ ($u^\omega$) and $V^*$ ($u^V$) are state (control input) vectors. The dimensions of $E$ ($u^E$), $P$ ($u^P$) and $V^*$ ($u^V$) are $M\times M$, while those of $Q$ ($u^Q$) and $\omega^*$ ($u^\omega$) are $m\times m$.Besides, dynamics of a BILPC are represented as
$$\begin{array}{l}
	\dot P_0 = u_0^P,\\
	\dot Q_0 = u_0^Q,\\
	\dot \omega _0^* = u_0^\omega,\\
	\dot V _0^* = u_0^V.
\end{array}$$
\par Thus, dynamics for AC and DC BESSs can be represented in a matrix form as 
\begin{equation}
	\left\{ \begin{array}{l}
		\dot x = Ax + u\\
		y = Bx
	\end{array} \right.
\end{equation}
where $x = col(E,P,\omega ^*,Q,V^*)$, $y =col(E,P,\omega ,Q,V)$, $u =col(u^E,u^P,u^\omega,u^Q,u^V)$,
$A = diag\left\{ {\left[ \begin{array}{*{20}{c}}
			0&\frac{1}{3600}I_M\\
			0&0
	\end{array} \right],0} \right\}$,
$B = diag\left\{ I_M,\left[ \begin{array}{*{20}{c}}
			B_{11}&0\\
			B_{13}&B_{14}
	\end{array}\right] \right\}$ with $B_{11} = \left[\begin{array}{*{20}{c}}
		I_M&{}\\
		-\left[I_m\quad \rm{0} \right]&I_m
\end{array} \right]$, ${B_{13}} = \left[ {\begin{array}{*{20}{c}}
		\rm{0}&\rm{0}\\
		diag\{0,-I_n\}&\rm{0}
\end{array}} \right]$ and $B_{14}=diag\{-I_m,0\}$. The dynamics for a BILPC can be represented in matrix form as 
\begin{equation}
	\left\{ \begin{array}{l}
		{{\dot x}_0} = {u_0}\\
		{y_0} = {B_0}x
	\end{array} \right.
\end{equation}
where ${x_0} =col(P_0,\omega_0^*,Q_0,V_0^*)$, ${u_0} =col(u_0^P,u_0^\omega ,u_0^Q,u_0^V)$, ${y_0} =col(P_0,\omega _0,Q_0,V_0)$ and $B_0= diag\{B_{01},B_{01}\}$ with $B_{01}=\left[ \begin{array}{*{20}{c}}
		1&0\\
		-1&1
\end{array} \right]$.
\subsection{Control objectives}
\par Generally, there are five common control objectives in BESSs, i.e., SoC balance, active/reactive power sharing and frequency/voltage restoration. Thus, given the characteristics of an ADHB, detailed control objectives in this paper can be expressed mathematically as follow. 
\\ (1) Active SoC balance should be achieved among all batteries both in AC and DC subgrids, i.e.,
$$\mathop {\lim }\limits_{t \to \infty } \left| {{E_i} - {E_j}} \right| = 0,i,j \in \{ 1,2, \cdots ,M\}.$$
\\(2) Active power sharing occurs among all BESSs both in AC and DC subgrids, while reactive power sharing should be reached within AC subgrid. That is,
$$\begin{array}{l}
	\mathop {\lim }\limits_{t \to \infty } \left| {K_i^P{P_i} - K_j^P{P_j}} \right| = 0,i,j \in \{ 1,2, \cdots ,M\},\\
	\mathop {\lim }\limits_{t \to \infty } \left| {K_i^Q{Q_i} - K_j^Q{Q_j}} \right| = 0,i,j \in \{ 1,2, \cdots ,m\}.
\end{array}$$
Actually, BILPC plays the role of energy conversion between two subgrids, i.e.,
$$\begin{array}{l}
	\mathop {\lim }\limits_{t \to \infty } {({P_0} - {\rm{(}}{P_L} - \sum\limits_{i = 1}^m {{P_i}} ))} = 0,\\
	\mathop {\lim }\limits_{t \to \infty } {({Q_0} - {\rm{(}}{Q_L} - \sum\limits_{i = 1}^m {{Q_i}} ))} = 0,
\end{array}$$
where ${P_L}$ and ${Q_L}$ are the total load in AC subgrid.
\\ (3) Frequency/voltage restoration (e.g., SC) should be fulfilled. Voltage at PCC of each DC BESS can achieve consensus. Frequency and voltage at PCC of each inverter and BILPC can achieve consensus and be restored to the reference, i.e.,
$$\begin{array}{l}
	\mathop {\lim }\limits_{t \to \infty } \left| {{V_i} - {V^r}} \right| = 0,i \in \{ 0,1,2, \cdots ,m\},\\
	\mathop {\lim }\limits_{t \to \infty } \left| {{V_i} - {V_j}} \right|=0,i,j \in \{ m + 1,2, \cdots ,M\},\\
	\mathop {\lim }\limits_{t \to \infty } \left| {{\omega _i} - {\omega ^r}} \right| = 0,i \in \{ 0,1,2, \cdots ,m\}.
\end{array}$$
\section{The proposed control framework}
\subsection{Graph theory}
\par For the MAS governing an ADHB, its communication network is described as a graph ${\cal G}({\cal V},{\cal E})$  where ${\cal V} = \left[ {{v_1},{v_2}, \cdots ,{v_M}} \right]$, namely vertex set, contains all agents corresponding to BESSs one by one. ${\cal E} \subseteq {\cal V} \times {\cal V}$ denotes the set of edges, which are communication links and do not have to be the same as electrical connections. Agent $j$ is called as a neighbor of agent $i$ if $i$ can access $j$, i.e. $({v_i},{v_j}) \in {\cal E}$, and the neighbor set of agent $i$ can be expressed as ${{\cal N}_i} = \{ ({v_i},{v_j}) \in {\cal E}|{v_j} \in {\cal V}\}$. The corresponding adjacency matrix is defined as ${\cal A} = [{a_{ij}}]$ and ${a_{ij}} = 1$ if $({v_i},{v_j}) \in {\cal E}$, which means agent $i$ can acquire the information of agent $j$, otherwise ${a_{ij}}=0$. The Laplacian matrix $L=\left[ {{l_{ij}}} \right]$ corresponding ${\cal G}$ is calculated by $l_{ii}=\sum\limits_{j \in {{\cal N}_i}} {{a_{ij}}}$ and $l_{ij}=-a_{ij}$. In particular, the pinning matrix $A_1= diag({a_i}) \subseteq {R^{M \times M}}$ represents whether it is connected with the virtual leader $\{ {\omega ^r},{V^r}\} $ and ${a_i}{\rm{ = 1}}$ if agent $i$ can access to the virtual leader, otherwise ${a_i}{\rm{ = 0}}$.
\subsection{The distributed control of a droop-controlled ADHB based on a cloud server}
\begin{figure} 
	\includegraphics[width=8cm]  {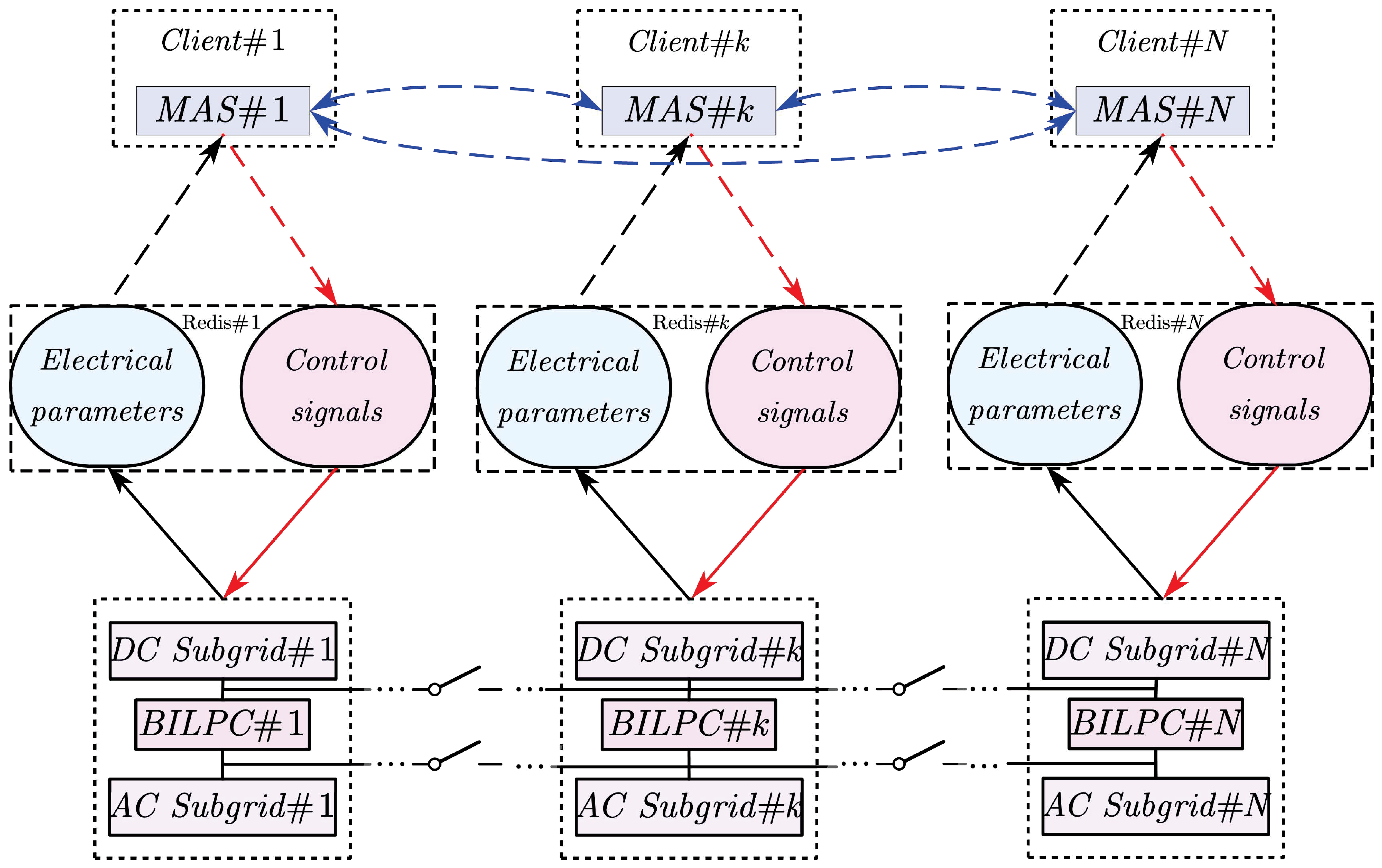}
	\caption{Framework of proposed distributed SC via cloud server.}\label{fig3}
\end{figure}
\par Similar to Internet of Things, each BESS is equipped with a control and communication agent in an EI. Agents in each group communicate with each other through local area networks (LANs) and can realize remote control of BESSs by accessing the Internet. Usually, these BESSs/ADHBs do not belong to the same stakeholder and their control signals are generated by MASs far from entity. Besides, if the number of agents/BESSs in each group can be modified online, it will be very convenient for remote control. Thus, a distributed, remote and flexible control framework is worth developing. Obviously, there is a set of parallel BILPCs in interconnected ADHBs in an EI paradigm. Fig. \ref{fig3} shows the working framework of remote distributed control of ADHBs in an EI based on a cloud server. 
\par The electrical parameters $\left\{E_i,P_i,Q_i\right\}$, $\left\{E_j,P_j\right\}$ and $\left\{P_0,Q_0\right\}$ collected in an ADHB are uploaded to the cloud server (such as Redis) by the simulator (such as OPA-RT). The control signals $\left\{\omega _i^*, V_i^*\right\}$, $\left\{V_j^*\right\}$ and $\left\{\omega _0^*, V_0^*\right\}$ are also downloaded locally from the cloud server via the simulator. The remote control scheme can adopt Google remote procedure call (gRPC) framework, which is composed of multiple clients and supporting servers. Agents communicate with each other in gRPC client/server manner and executes the SC algorithm. Each agent is both a server waiting for incoming data and a client of a neighbor server. The gRPC client/server accesses the cloud server through the Internet and downloads relevant data. The gRPC client/server is connected to the LAN through the switch to realize the data exchange between neighbors. After executing the relevant algorithms, the control signal is obtained and uploaded to the cloud server. The local controller updates the electrical parameters using the control signals downloaded from the cloud server by the simulator. Therefore, each agent obtains local information from the cloud server through the Internet and neighbor information through the LANs to realize remote control.
\subsection{The consensus controller for AC and DC subgrids design based on the Leader-Follower model}
\par As an efficient power structure in the community, although authors have designed a distributed SC solution in [29], the scheme for restoring voltage and frequency to the rated value of the utility grid is not given. Moreover, the frequency/voltage regulation scheme of a BILPC is not addressed. In the communication scheme, the agent managing BILPC needs to communicate with the agent managing AC BESS and the agent managing DC BESS. Thus, with the support of a cloud server, this paper will design a distributed scheme for the whole hybrid network based on MASs. In the context of a large-scale ADHB, considering the flexiblity scalability of the system in an EI paradigm, a power sharing controller for multiple parallel BILPC is designed. 
\par Based on MASs, distributed controllers for an ADHB are given in compact forms in (\ref{6a})-(\ref{6e}):
\begin{subequations}
	\begin{equation}
		\label{6a}
		u^E=-C^EL^0E
	\end{equation}
	\begin{equation}
		\label{6b}
	    u^P=-C^PL^0P
    \end{equation}
	\begin{equation}
		\label{6c}
	    u^\omega=-C^\omega(L^A\omega^*+A_1(\omega^*-P_A-\omega ^r))
    \end{equation}
    \begin{equation}
    	\label{6d}
	    u^Q=-C^QL^AQ
    \end{equation}
    \begin{equation}
    	\label{6e}
	    u^V=-C^V(LV^*+A_1(V^*-Q-V^r))
    \end{equation}
\end{subequations}
where $u^E$, $u^P$, $u^\omega$, $u^Q$ and $u^V$ are control input vectors, $C^E$, $C^P$, $C^\omega$, $C^Q$ and $C^V$ are coefficient matrices, $L$, $L^0$ and $L^A$ are Laplace matrices corresponding to communication topology and its sub-topologies, $A_1$ is the pinning matrix.
\par Thus, all control inputs can be synthesized into a matrix form
\begin{equation}
	u =  - C[{\cal L}x + {{\cal A}_0}(Bx - {x^r})]
\end{equation}
where all matrices involved are expressed by $x^r= col(0,0,\omega ^r,0,V^r)$, $C = diag\{C^E,C^P,C^\omega,C^Q,C^V\} $, ${\cal L} = diag\{ L^0,L^0,L^A,L^A,L\} $ and ${\cal A}_0 = diag\{0,0,A_1,0,diag\left\{ A_1,0 \right\}\} $.
\par Next, we consider a large-scale ADHB in an EI paradigm governed by some MASs. Without loss of generality, the same configuration is equipped in each ADHB. The number of agents governing all BESSs in each LAN is the same. Referring to the control framework in Fig. \ref{fig2}, we give a feedback control law as shown in (\ref{8}), 
\begin{equation}
	\label{8}
	\begin{array}{l}
		\left[ {\begin{array}{*{20}{c}}
				{{u_1}}\\
				{{u_2}}\\
				\vdots \\
				{{u_N}}
		\end{array}} \right] =  - \underbrace {\left[ {\begin{array}{*{20}{c}}
					{C{\cal L}}&0& \cdots &0\\
					0&{C{\cal L}}& \cdots &0\\
					\vdots & \vdots & \ddots & \vdots \\
					0&0& \cdots &{C{\cal L}}
			\end{array}} \right]\left[ {\begin{array}{*{20}{c}}
					{{x_1}}\\
					{{x_2}}\\
					\vdots \\
					{{x_N}}
			\end{array}} \right]}_{{\rm{Intra - MAS}}}\\
		{\rm{            }} - \underbrace {\left[ {\begin{array}{*{20}{c}}
					{{{\tilde l}_{11}}C\Delta }&{{{\tilde l}_{12}}C\Delta }& \cdots &{{{\tilde l}_{1N}}C\Delta }\\
					{{{\tilde l}_{21}}C\Delta }&{{{\tilde l}_{22}}C\Delta }& \cdots &{{{\tilde l}_{2N}}C\Delta }\\
					\vdots & \vdots & \ddots & \vdots \\
					{{{\tilde l}_{N1}}C\Delta }&{{{\tilde l}_{N2}}C\Delta }& \cdots &{{{\tilde l}_{NN}}C\Delta }
			\end{array}} \right]\left[ {\begin{array}{*{20}{c}}
					{B{x_1}}\\
					{B{x_2}}\\
					\vdots \\
					{B{x_N}}
			\end{array}} \right]}_{{\rm{Inter - MAS}}}\\
		{\rm{            }} - \underbrace {\left[ {\begin{array}{*{20}{c}}
					{C{{\cal A}_0}}&0& \cdots &0\\
					0&{C{{\cal A}_0}}& \cdots &0\\
					\vdots & \vdots & \ddots & \vdots \\
					0&0& \cdots &{C{{\cal A}_0}}
			\end{array}} \right]\left[ {\begin{array}{*{20}{c}}
					{B{x_1} - {x^r}}\\
					{B{x_2} - {x^r}}\\
					\vdots \\
					{B{x_N} - {x^r}}
			\end{array}} \right]}_{{\rm{Pinning\,Nodes}}}
	\end{array}
\end{equation}
which contains intra-group, inter-group interactions and pinning effect, where the Laplacian matrix $\tilde L = \left[ {{{\tilde l}_{ij}}} \right] \in R{^{N \times N}}$ denotes the interaction among groups. The matrix $\Delta $ is used to select an agent in each group which is in charge of external communication and data exchange. Without losing generality, the control gains of the interaction term are selected as $C$ and $\Delta={\cal A}_0$. Thus, (\ref{8}) can be rewritten in compact form as
\begin{equation}
	\begin{array}{l}
		U =  - ({I_N} \otimes C{\cal L} + \tilde L \otimes {\cal A}_0 B)X - ({I_N} \otimes C{{\cal A}_0})\\
		\qquad \, \times (({I_N} \otimes B)X - {X^r})
	\end{array}
\end{equation}
where $\otimes$ denotes Kronecker product. Thus, the closed-loop system of a large-scale ADHB is represented as
\begin{equation}
	\begin{array}{l}
		\dot X = ({I_N} \otimes A)X - ({I_N} \otimes C)(({I_N} \otimes {\cal L} + \tilde L \otimes {\cal A}_0 B)X\\
		\qquad \ - ({I_N} \otimes {{\cal A}_0})(({I_N} \otimes B)X - {X^r})
	\end{array}
\end{equation}
\\ Define the state error $\delta  = ({I_N} \otimes B)X - {X^r}$. Thus, the corresponding error system can be derived as
\begin{equation}
	\begin{array}{l}
		\dot \delta  = ({I_N} \otimes B)\dot X\\
		\; \; \; =  - ({I_N} \otimes B)({I_N} \otimes C)(({I_N} \otimes {\cal L} + \tilde L \otimes {\cal A}_0 B)X\\
		\qquad + ({I_N} \otimes {{\cal A}_0})\delta )\\
		\; \; \; =  -({I_N} \otimes C)({I_N} \otimes B{\cal L} + \tilde L \otimes  B{\cal A}_0+ {I_N} \otimes {{B\cal A}_0})\delta\\
		\; \; \; =  - \Phi \delta
	\end{array}\label{11}
\end{equation}
Select the Lyapunov function $V = {\delta ^T}\delta $ for the error system (\ref{11}), find its derivative as follow
\begin{equation}
	\dot V  =- {\delta ^T}(\Phi^T+\Phi) \delta
\end{equation}
\par Therefore, only if $(\Phi^T+\Phi)$ is non-negative definite, the system stability criterion can be assured, i.e. $\dot V \le {\rm{0}}$. It is easy to verify $(B+B^T){{\cal A}_0}$ and $(B{\cal L}+{\cal L}B^T){{\cal A}_0}$ are non-negative. It can be concluded with the above analysis, $({I_N} \otimes C) > 0$, ${\cal L} \ge 0$ and $\tilde L \ge 0$ that $(\Phi^T+\Phi)\ge {\rm{0}}$. Thus, the error system in (\ref{11}) is asymptotically stable.
\par \textit{Remark 1}: Compared with [30], in which the authors designed the SC scheme by the measured power, frequency and voltage, the scheme designed in this paper only needs to use the measured power. The regulated no-load frequency and voltage signals are sent to the local controller and directly stored in cloud server.
\par \textit{Remark 2}: In order to illustrate the progressiveness of the designed controller, we compared it with the common controller constructed with the frequency and voltage measurements shown below.
	$$\begin{array}{l}
		U =  - ({I_N} \otimes C)({I_N} \otimes {\cal L} + \tilde L \otimes \Delta)X\\
		\quad \quad- ({I_N} \otimes {{\cal A}_0}) (X - {X^r}))
	\end{array}$$
	,where $X$ is a column vector composed of measured values. Define $\delta_1=X-X^r$ and 
	$$\begin{array}{l}
		\dot \delta_1  =-\Phi_0 \delta=  - ({I_N} \otimes C)({I_N} \otimes {\cal L} + \tilde L \otimes {\cal A}_0+{I_N} \otimes {{\cal A}_0})\delta
	\end{array}$$. It is easy to conclude that the matrix has the same eigenvalues as $\Phi$ and $\Phi_0$. Thus, the performance of our SC will not deteriorate due to use the nominal frequency/voltage. 
\subsection{Secondary controller for BILPC}
\par For a BILPC, as a vassal of the AC subgrid, the outpower should be regulated in order to keep balance between load and outpower. 
\par Based on the above analysis, controllers for active/reactive power, frequency and voltage of BILPC are designed as follow. 
\begin{subequations}
	\begin{equation}
		u_0^P{\rm{ = }} - C_0^P({P_0} - P_0^r)
	\end{equation}
    \begin{equation}
    	u_0^\omega {\rm{ = }} - C_0^\omega (\omega _0^* - {P_0} - {A_0}({\omega ^*} - {P_A}))
    \end{equation}
    \begin{equation}
    	u_0^Q{\rm{ = }} - C_0^Q({Q_0} - Q_0^r)
    \end{equation}
    \begin{equation}
    	u_0^V{\rm{ = }} - C_0^V(V_0^* - {Q_0} - {A_0}({V^*} - Q))
    \end{equation}
\end{subequations}
where ${A_0} = {[{a_{0i}}]_{1 \times m}}$, if  ${a_{0i}} = 1$, the virtual nominal frequency and voltage of BILPC are regulated by the states of BESS $i$, $P_0^r$ and $Q_0^r$ are the reference power for BILPC.

\par Then, above controllers can be rewritten the matrix form as follow
\begin{equation}
	{u_0} =  - {C_0}({B_0}{x_0} - x_0^r)
\end{equation}
where $x_0^r ={\cal A}_1 B_1x+col(P_0^r,0,Q_0^r,0)$, and the correlation matrices are $C_0 = diag\{C_0^P,C_0^\omega ,C_0^Q,C_0^V\}$, ${\cal A}_1 = diag\{ 0,A_0,0,[A_0,0]\}$, $B_1 = \left[ {\begin{array}{*{20}{c}}
		0&{\left[ {\begin{array}{*{20}{c}}
					B_{11}&0\\
					B_{13}&B_{14}
			\end{array}} \right]}
\end{array}} \right]$. Thus, controllers of a set of BILPCs in parallel can be represented in compact form, as
\begin{equation}
		   \begin{array}{*{20}{l}}
			{{U_0} =- ({I_N} \otimes {C_0})(({I_N} \otimes {B_0}){X_0} -  X_0^r)}
		\end{array}
\end{equation}
, where $X=col(x_{01},x_{02},\cdots,x_{0N})$ and $X^r=col(x^r_{01},x^r_{02},\cdots,x^r_{0N})$.
\par \textit{Remark 3}: It can be seen from Fig. 4 that the BILPC of each ADHB is only regulated by the relevant states of an AC BESS. Under the plug-and-play framework, control objectives can be achieved without communication among BILPCs in an EI. That is, the control law (15) is decentralized.
\par \textit{Remark 4}: To calculate the reference power $P_0^r$ and $Q_0^r$, the load power of each bus is needed. In this paper, suppose that there is a scheme in the cloud server that can well predict the load power as shown in [31] and [32]. The accuracy of load prediction of each bus will not affect the frequency/voltage convergence. But obviously inaccurate prediction will have a great impact on the physical network. If the predicted value is not accurate, there will be obvious imbalance between supply and demand in the physical network.
\subsection{Communication delay in controllers}
\par It is assumed that a uniform delay $\tau $ in each communication link of integrator MASs, and $\tau  < {\tau _m}$. In [33], authors give a function (16) of the largest eigenvalue of the Laplacian matrix to estimate the upper boundary of the uniform communication delay
\begin{equation}
	{\tau _m} =\pi/{2\lambda_{max}(\Phi)}
\end{equation}
\par It is worth mentioning that for the given communication topology and control gain $C$, there is always a maximum time delay ${\tau _m}$ that can guarantee the stability of the system. Likewise, for the given communication topology and its maximum time delay ${\tau _m}$, the system stability can be kept by selecting proper control gain matrix $C$. Based on [20], it can be concluded that the smaller the gain, the greater the maximum communication delay. 
\begin{figure} 
	\includegraphics[width=8cm]  {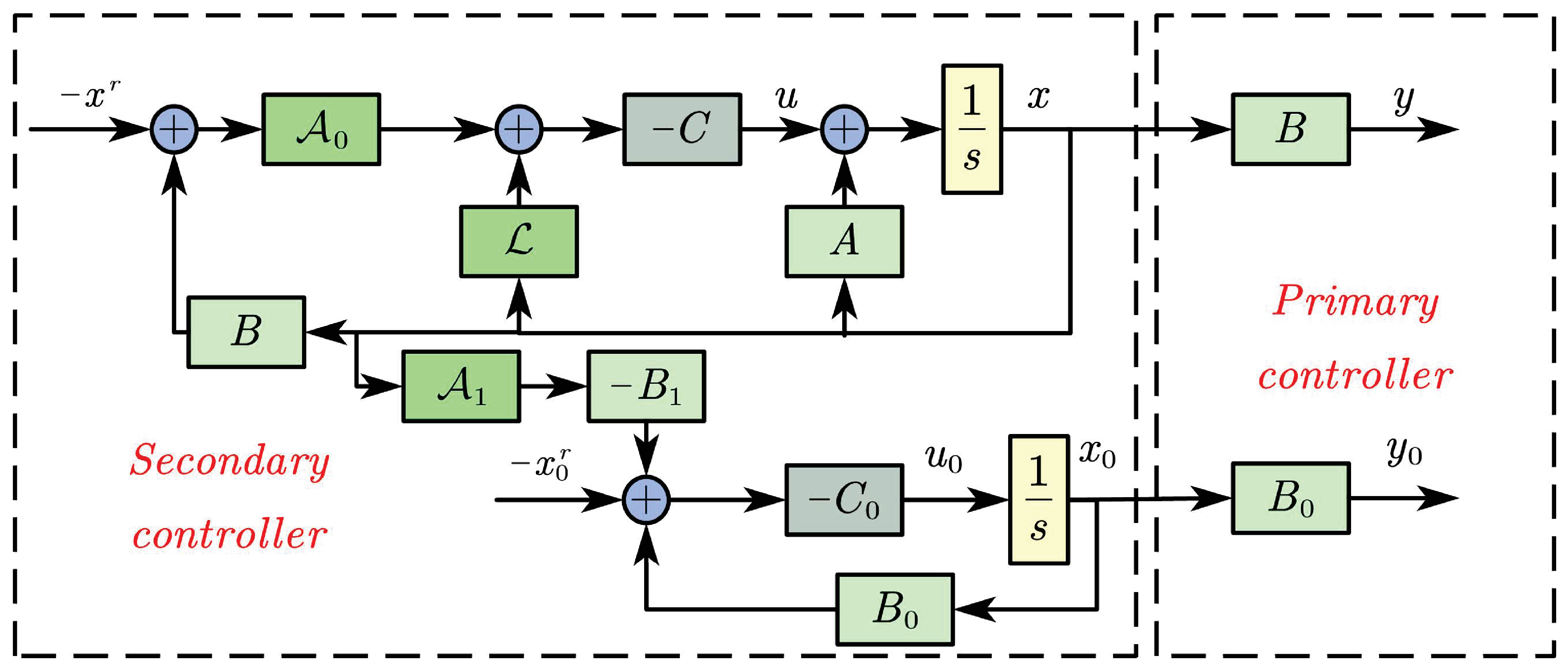} 
	\caption{PC and designed SC.}\label{fig4}
\end{figure}
\par \textit{Remark 5}: In the smart grid, communication delays are common in monitoring and control systems of each application. However, different applications have different requirements for time delay. For example, the tolerable maximum communication delays are 100 ms and 200 ms in the substation automation and the wide-area situational awareness systems respectively [34]. In a distribution automation system, the tolerable maximum communication delay, as defined by the IEEE and international electro-technical commission (IEC), is 15 ms to tolerate any potential destabilizing factor [34]. According to the Geschgorin Disk Estimator, ${\lambda _{\max }}(\Phi ) \le 2(D(\Phi ) - 1)$, where $D(\Phi )$ is the dimension of the matrix $\Phi $. Combined with (17), if $D(\Phi ) < 54$, ${\tau _m}>15 ms$, and the system is stable within the communication delay $15{\rm{ ms}}$.

\section{Case study}
\begin{figure}
	\centering
	\subfigure[Single line diagram of an ADHB.]{\includegraphics[width=0.6\hsize, height=0.4\hsize]{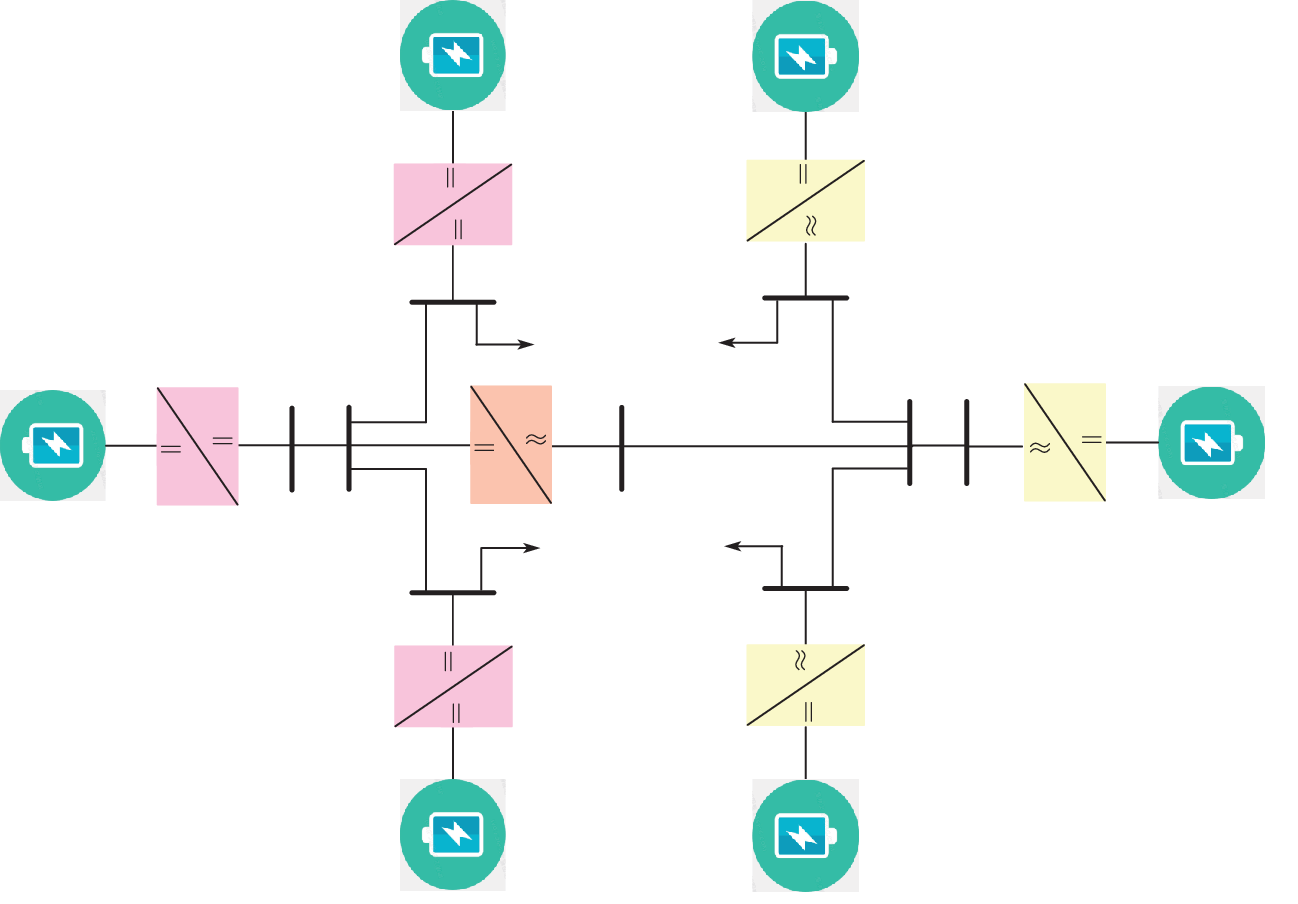}\label{fig4a}}\hspace{1cm}
	\subfigure[Communication graph for the ADHB in (a).]{\includegraphics[width=0.7\hsize, height=0.3\hsize]{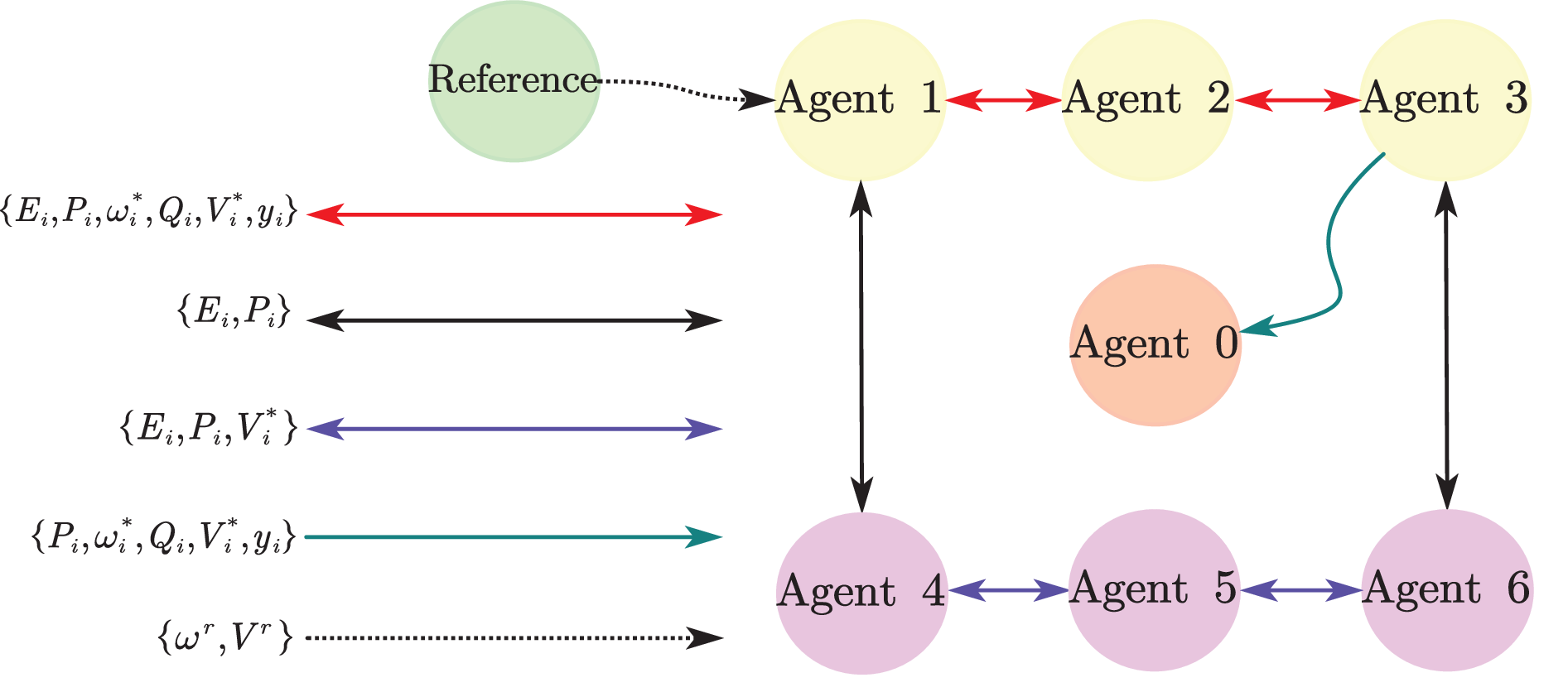}\label{fig4b}}
	\caption{Single line diagram of an ADHB and its communication
		graph.}
	\label{fig4}
\end{figure}
\par To verify the proposed control framework, four test cases are considered in this section. And an ADHB, which contains 3 BESSs in AC subgrid and 3 BESSs in DC subgrid, and its supporting cloud service communication network are shown in Fig. \ref{fig4}, which are used in Case 1-3 and extended in Case 4. Thus, the four test cases are organized as follow.
\par In Case 1-3, an isolated ADHB containing 3 AC BESSs, 3 DC BESSs and a BILPC is adopted to validate the proposed control framework. The three cases are conducted under ideal condition, communication failures and uniform communication delays.
\par In Case 4, three ADHBs with the same configuration are interconnected in the simulation to validate the group play-and-plug of the proposed control framework.
\subsection{Case 1: Validation of the Proposed Control Framework}
\par This case is arranged to verify the proposed control scheme for an ADHB. To facilitate the analysis of simulation results, this case and following cases are simulated in per unit system. In order to test the performance of the controller when the load changes, active and reactive loads increase by 0.5 $p.u.$ at $t = 50$ and $t = 80$ respectively. Fig. 6 and 7 are the simulation results, where all states of each BESS are shown in Fig. 6, and the states of BILPC are shown in Fig. 7.
\begin{figure}
	\includegraphics[width=8cm]{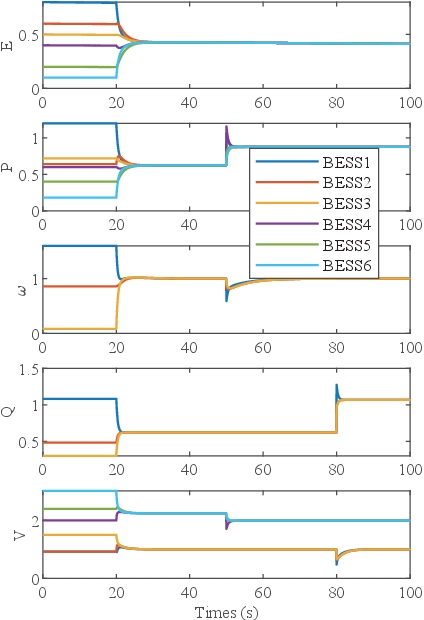} \\
	\caption{ All states of each BESS in Case 1.} 
\end{figure}
\begin{figure}
	\includegraphics[width=8cm]{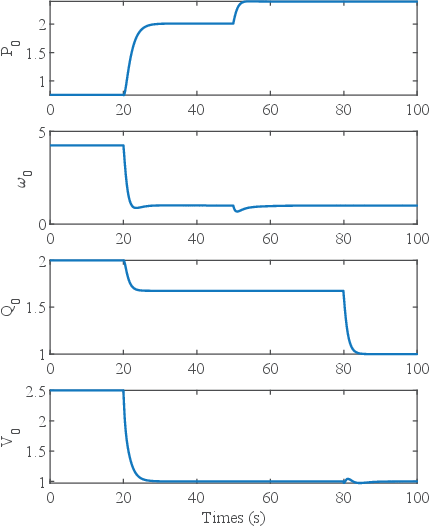}\\
	\caption{All states of BILPC in Case 1.} 
\end{figure}
\par It can be observed easily in Fig. 6 that active balance of SoC, active/reactive power sharing and frequency/voltage consensus can be well kept effectively even if the load changes. For the AC subgrid, frequency/voltage can be restored to the reference. In Fig. 7, frequency and voltage of the BILPC can also restore to the reference value. And, output power reach stable. The results of this case fully prove that the distributed control method proposed for the remote ADHB is valid. The next simulation and its results will be compared with Case 1.
\subsection{Case 2: Communication Failures}
\par Case 2 is designed to test the influence of communication failure on the proposed control scheme. The simulation events for this case are explained in detail in Fig.8.
\begin{figure}
	\includegraphics[width=8cm]{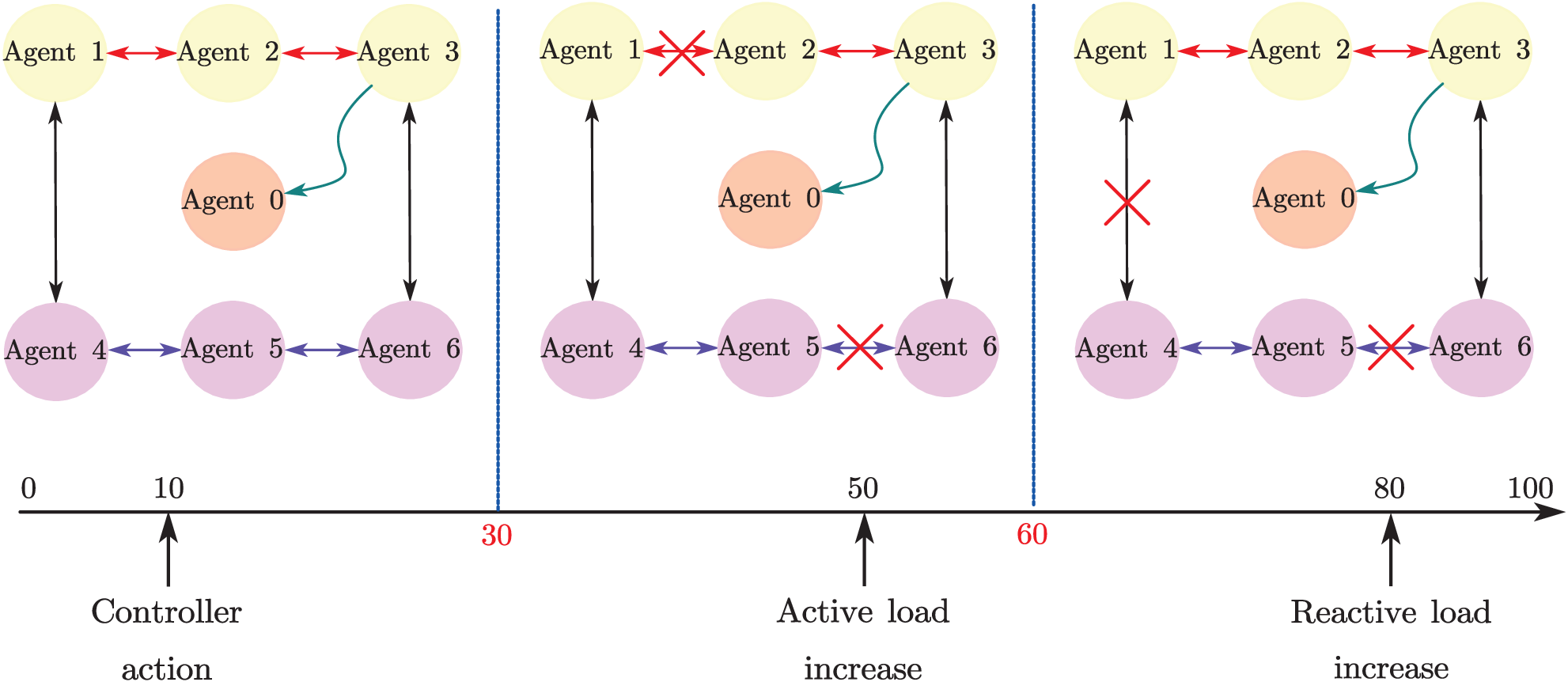}
	\caption{Simulation events in Case 2.} 
\end{figure}
\begin{figure}
	\includegraphics[width=8cm]{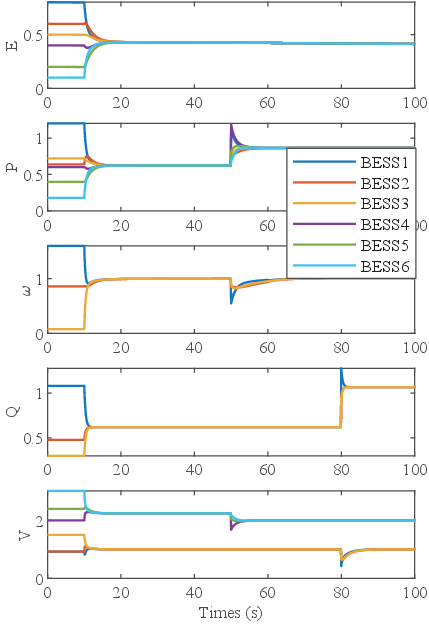} \\
	\caption{All states of each BESS in Case 2.} 
\end{figure} 
\begin{figure}
	\includegraphics[width=8cm]{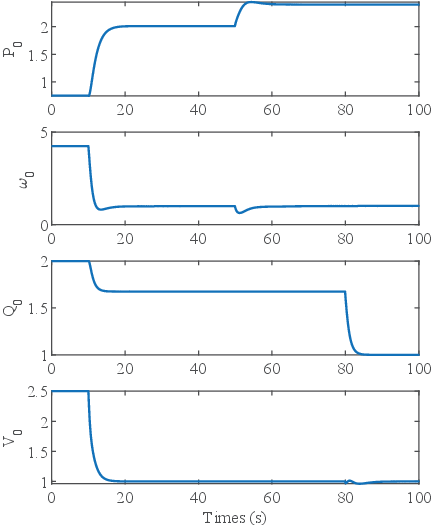}\\
	\caption{All states of BILPC in Case 2.} 
\end{figure}
\par The controller acts at 10s. Before 30s, any communication failure did not occur. It is assumed that there exists communication failures on link 1-2 and 5-6 from 30s to 60s, while communication failures on link 1-4 exist from 60s to 100s. Load switch and controller parameters are the same as Case 1. The simulation results are shown in Fig. 9 and 10. Compared with Case 1,  there is not a significant impact on each BESS and the BILPC under communication failures.
\subsection{Case 3: Communication Delays}
\begin{figure}
	\includegraphics[width=8cm]{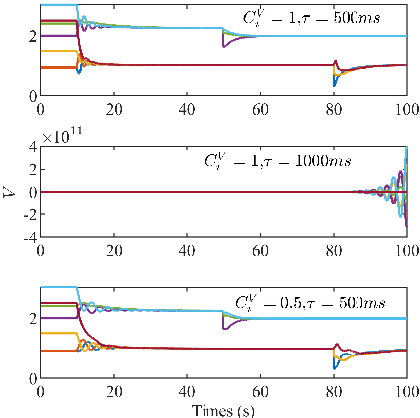}\\
	\caption{Voltage of each BESS and BILPC with communication delays in Case 3.} 
\end{figure}
\par In this case, the performance of the control framework with an uniform communication delay are test here. Firstly, a simulation with delay 500 ms is executed, and PCC voltage of each BESS and the BILPC are shown in the first subgraph of Fig.11. Compared with Fig. 6 and 7, the existence of communication delay makes the voltage curve produce local small oscillation and slows down the convergence speed. When the delay increases to 1000 ms, as shown in the second subgraph, voltage becomes divergent. This ADHB system becomes unstable. By reducing control gains, as shown in the third subgraph of Fig.11, the system restores to be  stable. However, the convergence speed becomes more slowly. It can be seen from this that under the effect of communication delay, there is a tradeoff between the convergence speed and the stability of this system.
\subsection{Case 4: Group Plug-and-Play}
\begin{figure}
	\includegraphics[width=8cm]{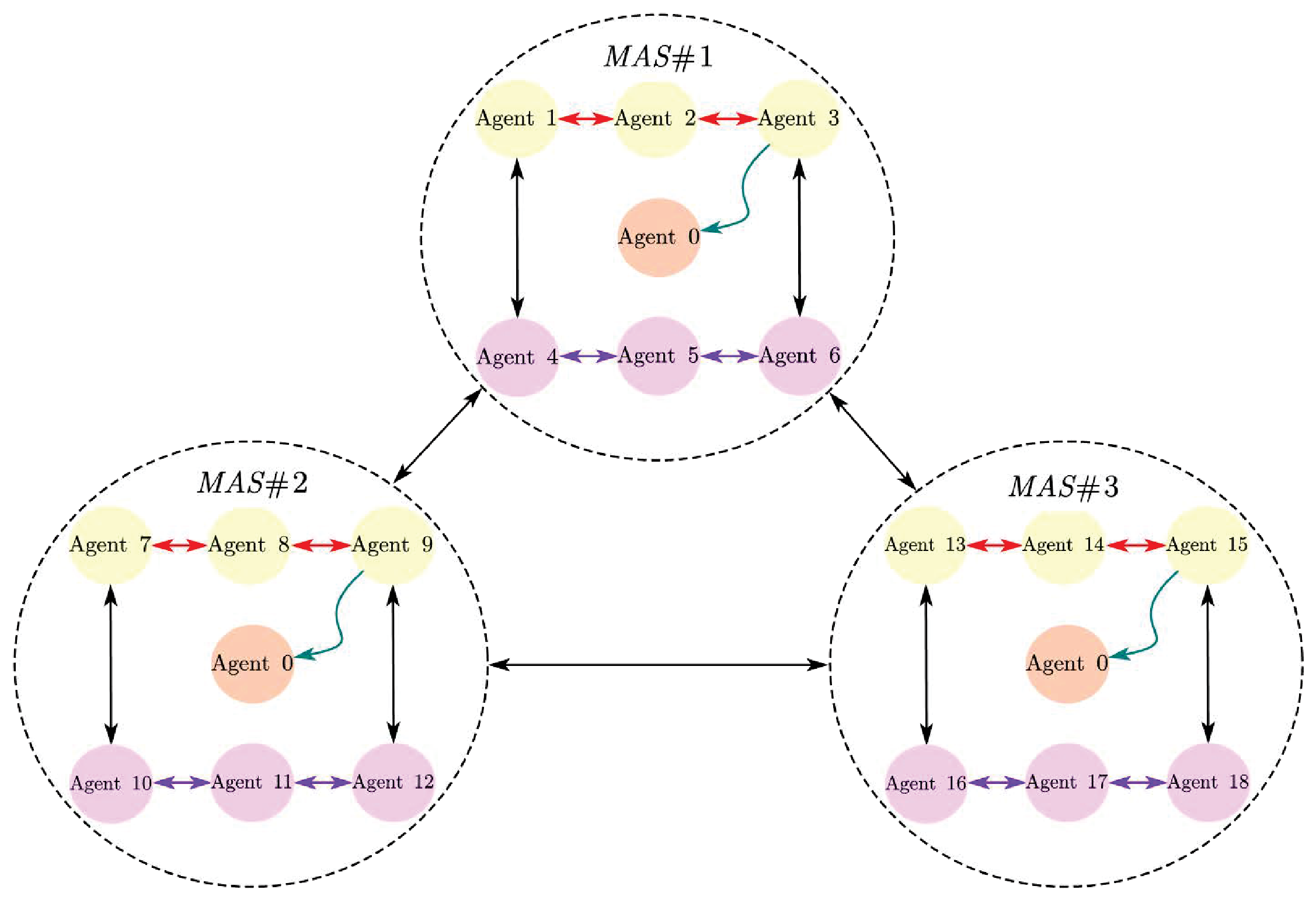}\\
	\caption{Communication graph among 21 BESSs managed by three group MASs in Case 4.} 
\end{figure}
\par In Case 4, the multiple ADHBs framework proposed is validated for group plug-and-play. Three ADHBs, with the same configuration except the load, governed by three groups MASs are adopted in this case as shown in Fig.12, in which interactions between any two groups can be managed by set up or remove a communication link. The simulation events are described in Fig. 13. The simulation results are shown in Fig. 14 and 15.
\par It can be observed obviously that SoC balance and proportional power sharing among any two group can be reached by adding a communication link. Besides, power sharing of parallel BILPCs also can be reached by adding a communication link. After removing the added communication link, power sharing between groups no longer exists, but within the group it is not affected.

\begin{figure}
	\includegraphics[width=8cm]{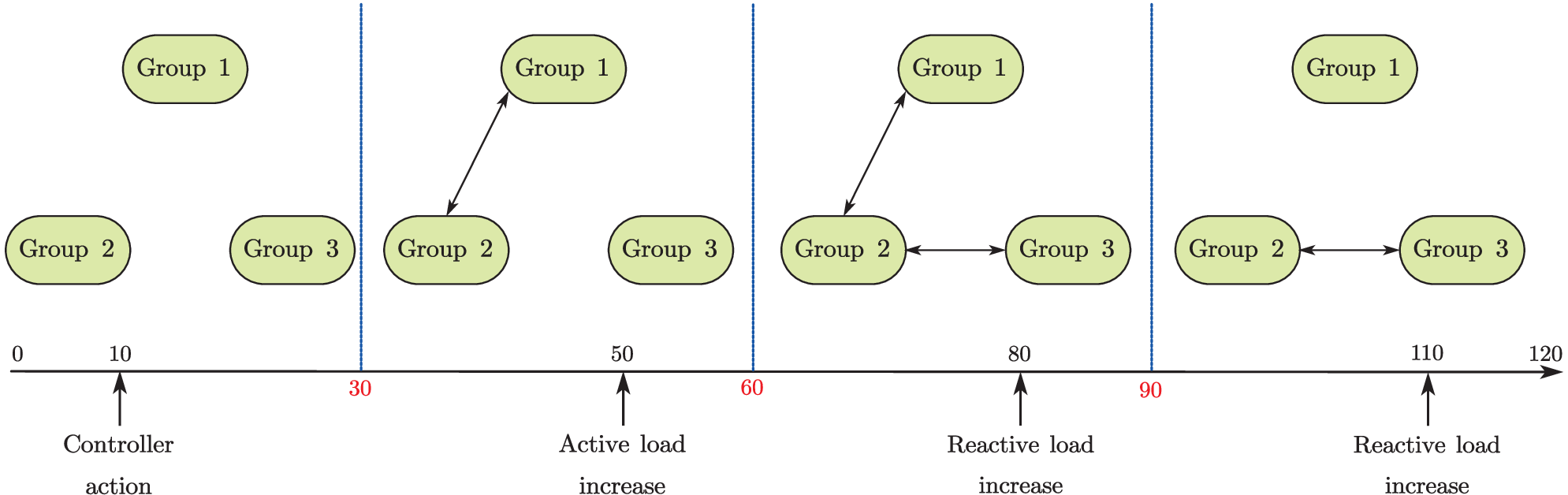}\\
	\caption{ Simulation events in Case 4.} 
\end{figure}
\begin{figure}
	\includegraphics[width=8cm]{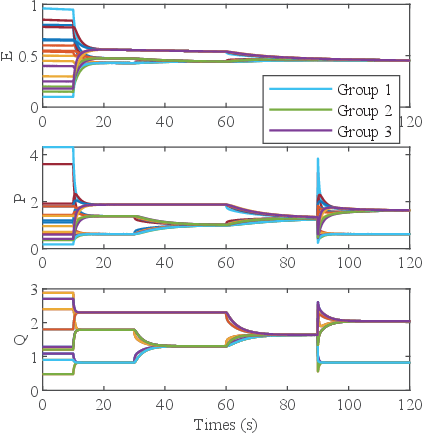}\\
	\caption{ SoC balance and power sharing among all BESSs in Case 4.} 
\end{figure}
\par Though the simulation results in Case 4, the following can be concluded that
\\(1)  The global consensus can be guaranteed by selected a fully connected communication graph. For a partially connected graph, if it can be divided to some fully connected subgraphs, local consensus within each group can be ensured.
\\(2) The proposed control framework supports the group plug-and-play by change the connectivity among groups. However, this operation does not change the intragroup communication graph.
\begin{figure}
	\includegraphics[width=8cm]{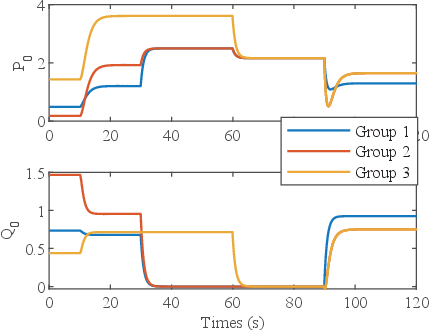}\\
	\caption{Power sharing among all BILPCs in parallel in Case 4.} 
\end{figure}
\section{Conclusion}
\par In this article, a distributed remote control framework is proposed for ADHBs in an EI by different MAS groups. In the designed control scheme, the transmission method of different data packets among agents governing AC and AC, AC and DC BESSs is defined. By arranging the inter-group communication link among MASs, flexible control of BESSs in an EI can be realized effectively, and testing of the influence of time delay on stability is considered. Firstly, the simulation results showed that the designed controller is effective under step load change and communication failures. Secondly, with the communication delay increasing from 500ms to 1000ms, the system transitions from stable to unstable. However, by reducing control gains, the system become stable again. Finally, the flexibility and scalability of the proposed approach is test by three ADHBs, governed by three groups MAS, interconnected in an EI. Future work will focus on the following two important aspects: (1) a distributed scheme for estimating the global power mismatch, especially under the time varying load; (2) a distributed scheme for power sharing among parallel BILPCs. 
\appendices

\ifCLASSOPTIONcaptionsoff
  \newpage
\fi

\vspace{-15 mm} 
\begin{IEEEbiography}[{\includegraphics[width=1in,height=1.25in,clip,keepaspectratio]{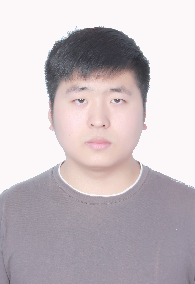}}]{Yalin Zhang}
received a B.E. degree and MA.Sc. degree in Automation and Control science and Engineering from Henan Polytechnic University, China, in 2016 and 2020, respectively. He is now pursuing a Ph.D degree in Control science and Engineering at Nankai University, China. His research interests include multi-agent system modeling and distributed control and optimization in energy internet.
\end{IEEEbiography}
\vspace{-10 mm} 
\begin{IEEEbiography}[{\includegraphics[width=1in,height=1.25in,clip,keepaspectratio]{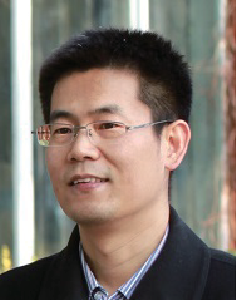}}]{Zhongxin Liu}
    received his B.S. degree in automation and the Ph.D. degree in control theory and control engineering from Nankai University, Tianjin, China, in1997 and 2002, respectively. \par He has been with Nankai University, where he is currently a Professor with the Department of Automation. His research interests include multi-agent systems, nonlinear control theory, networked control system, as well as complex network theory and its application.
\end{IEEEbiography}
\vspace{-10 mm} 

\begin{IEEEbiography}[{\includegraphics[width=1in,height=1.25in,clip,keepaspectratio]{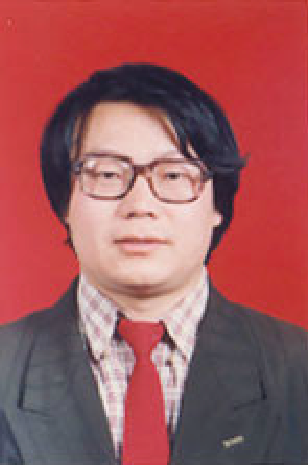}}]{Zengqiang Chen}
	 received his B.S. degree in mathematics and the M.S. and Ph.D. degrees in control theory and control engineering from Nankai University, Tianjin, China, in 1987, 1990, and 1997, respectively. 
	 \par He has been with Nankai University, where he is currently a Professor with the Department of Automation. His research interests include predictive control technology, complex network system, chaotic system theory and its application in information security.
\end{IEEEbiography}
\end{document}